\documentclass{aa}
\usepackage{graphics}
\begin{document}
\thesaurus{06(08.02.3; 08.03.1; 08.16.4; 09.09.2 V Hya)}
\title{The light curve and evolutionary status of 
the carbon star V Hya}

\author{G. R.\, Knapp\inst{1}, S. I.\, Dobrovolsky\inst{1}
\and Z. Ivezi\'c\inst{1},
K. Young\inst{2} \and M. Crosas\inst{2}, J. A.\, Mattei\inst{3}
\and  M. P. \,Rupen\inst{4}}

\offprints{G. R. Knapp}

\institute{Department of Astrophysical Sciences, Princeton University,
Princeton, NJ 08544, USA; gk, sid, ivezic@astro.princeton.edu
\and Harvard-Smithsonian Center for Astrophysics, 60 
Garden St., Cambridge, MA 02138, USA; rtm, mcrosas@dolson.harvard.edu
\and American Association of Variable Star Observers, 25 Birch St.,
Cambridge, MA 02128-1205, USA; jmattei@aavso.org
\and National Radio Astronomy Observatory, P.O. Box 0, 1003 Lopezville Rd.,
Socorro, NM 87801, USA; mrupen@nrao.edu}

\date{Received 15 December 1998 / Accepted 1 July 1999}
 
\maketitle

\begin{abstract}

V Hya, an evolved carbon star with a complex circumstellar
envelope, has two variability periods, $\rm 530^d$ and $\rm 6000^d$
(17 years).  We analyze recent light curve data and show that both 
variations have been present for at least 100 years and have been 
regular over this time.  The $\rm 530^d$ period and its
$\rm 1.5^m - 2^m$
amplitude show that V Hya is a Mira variable.  We suggest that 
the star is in a binary system (as also suspected from the structure
of the circumstellar envelope) and that the 17-year variation is
due to extinction by circumstellar dust orbiting with the companion.
The properties of the envelope found from molecular line observations:
the fast molecular wind, the relatively small size of the dense 
circumstellar envelope, and the high mass loss rate, all suggest that
V Hya has entered its `superwind' phase. However, its spectral type,
period, colors, and lack of ionizing radiation
show that the star is still on the
AGB.  These properties add to the evidence that the complex structures
of many planetary nebulae, including fast stellar winds, originate 
during the final phases of mass loss on the AGB.


\end{abstract}

\section{Introduction}

V Hya is a cool evolved N type carbon star with several unusual 
properties, including:
(1) two variability periods, of about 530 and 6000 days (Mayall 
\cite{mayall});
(2) possible shock-excited forbidden-line emission similar to
that seen from Herbig-Haro objects (Lloyd Evans \cite{lloyd});
(3) continuum emission at radio wavelengths far in excess of the
expected photospheric flux (Luttermoser \& Brown \cite{lb});
(4) a circumstellar envelope which has a flattened or disk-like
shape (Tsuji et al. \cite{tsuji}; Kahane et al. \cite{kahane88};
Kahane et al. \cite{kahane96}); and
(5) a 200 $\rm km~s^{-1}$, possibly bipolar,
wind seen in CO(v = 0--1) absorption (Sahai \& Wannier \cite{sahai88}),
in KI absorption and emission (Plez \& Lambert \cite{plez}),
in optical forbidden-line emission (Lloyd Evans \cite{lloyd}) and in CO
millimeter-wavelength spectral line emission (Knapp et al. \cite{knapp}, 
hereafter Paper I). Because V Hya is
one of only two known evolved stars with a fast molecular wind which
still has the infrared colors of an AGB star (the other being $\rm \Pi^1$
Gru), it may be in the very earliest stages of 
evolution away from the AGB.

The present paper discusses several new observations of V Hya and
its envelope, which we integrate with previous observational
results to learn about the
evolutionary status of V Hya and the behavior of stars as they
evolve away from the AGB.  The main part of the paper (Sect. 2) 
is an analysis of archival observations
of the star's light curve made between October 1961 and July 1996.
Appendix A summarizes
new observations of the star's radio frequency emission and 
of the molecular circumstellar envelope. 
Section 3 discusses the implications of these and previous
data for the evolutionary status of V Hya, and
gives the conclusions.

\begin{table}
\caption[]{Basic Properties of V Hya}
\vspace{0.5cm}
\begin{tabular}{llllll}
Spectral Type& N:C7,5\cr
Variable Type& Mira\cr
Distance& 500 pc\cr
$\rm T_{eff}$& 2650 K\cr
$\rm L_{bol}$& $\rm 1.4 \times 10^4 ~L_{\odot}$\cr
$\rm R_{\star}$& $\rm 3.8 \times 10^{13}~cm$\cr
$\rm P_1$: period& $\rm 6160^d \pm 400^d$\cr
$\rm P_1$: amplitude& $\rm 3.5^m$\cr
$\rm P_1$: maximum& JD2446937\cr
$\rm P_1$: minimum& JD2450017\cr
$\rm P_2$: period& $\rm 529.4^d \pm 30^d$\cr
$\rm P_2$: amplitude& $\rm 1.5^m$\cr
$\rm P_2$: maximum& JD2450017\cr
$\rm P_2$: minimum& JD2449617\cr
\end{tabular}
\end{table}

Basic data for V Hya, including properties derived in this paper,
are summarized in Table 1.
Recently, Hipparcos (Perryman et al. 
\cite{perryman}) has provided accurate astrometric data for V Hya.
The parallax of the star is too small to measure ($\rm \pi ~ = ~ 0.16
~ \pm ~ 1.29$ milliarcseconds) giving a 2$\sigma$ lower limit to the
distance of 400 pc. Bergeat et al. (\cite{bergeat})
discuss the period-luminosity relationship of carbon
variables using Hipparcos data, and suggest an absolute K magnitude 
of $\rm -9.05^m$ for 
V Hya based on its variability period of 530 days,  giving a distance
of 550 pc.  We assume a round-number distance of 500 pc in this paper.

\section{The Light Curve of V Hya}

Mayall (\cite{mayall}) analyzed data compiled by the American Association of
Variable Star Observers (AAVSO) between 1884 and 1965.  These data show a more
or less regular variability with a period of 533 days and
a peak-to-peak variation of $\rm \sim 2^m$ at V, plus a longer-term variation
every 6500 days (17 - 18 years) with deep minima ($\rm \sim 5^m - 6^m$).
Mayall's data also show some evidence for a third, much shorter, period,
$\rm 30^d$. Kholopov et al. (1985), in the notes to Volume 2 of the `Catalogue
of Variable Stars' note that V Hya has two superposed variations, a 
long-period variation with P = $\rm 6670^d$ and amplitude $\rm 3.5^m$,
and a $\rm 531^d$ period with amplitude $\rm 1.1^m$ to $\rm 2^m$.
The $\rm 530^d$ period is typical for stars at the tip of the
AGB, but the 17 year period, and the depth of the minimum, is very
unusual, not exhibited by any other AGB star (some supergiants,
e.g. $\alpha$ Ori and $\mu$ Cep, have both long and short periods, though
with much smaller amplitudes). V Hya is classified 
as a semi-regular (SRa) variable (Kholopov et al. \cite{kholopov}), but
as shown, for example, by Kerschbaum (\cite{kerschbaum}), the SRa class
contains a mixture of Miras and ``true'' SR (SRb) variables. 
We suggest below that V Hya is a $\rm 530^d$ Mira variable, and that 
the $\rm 6000^d$ period is unrelated to pulsation.

\begin{figure}
\resizebox{\hsize}{!}{\includegraphics{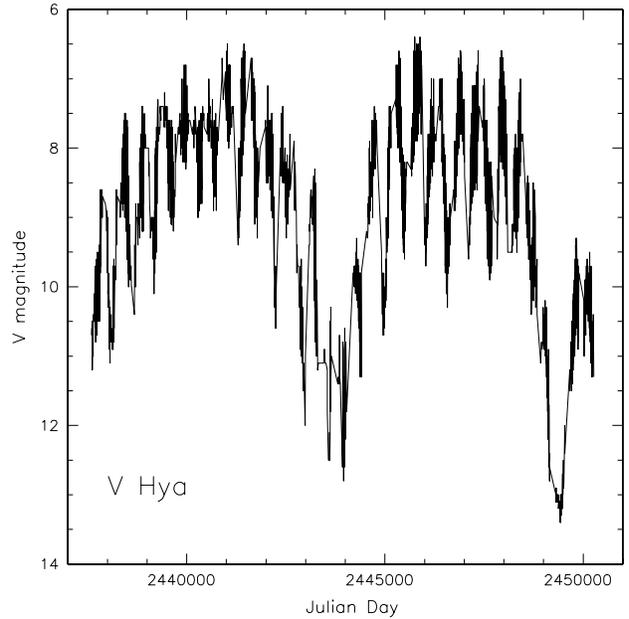}}
\caption{V band light curve of V Hya between October 1961 and July
1996 from the AAVSO archives.}
\label{lightcurve}
\end{figure}

We obtained the more recent light curve data for V Hya from the AAVSO
archives.  These data run from JD 2437602 (October 29, 1961, with a small
overlap with the data discussed by Mayall (\cite{mayall}) to JD 2450268 (July 3 1996).
They are plotted in Fig.\ref{lightcurve}, 
and the light curve shows a continuation
of the behavior observed by Mayall (\cite{mayall}); periodic variability at
$\rm 530^d$, and three deep minima.  

The total time interval covered by these observation is 12666 days,
and there are 3260 data points.  Were the sampling regular, the 
interval would be about 3.9 days, and the shortest period that could be
found, corresponding to the Nyquist frequency, would be 7.8 days.

The irregular sampling of the light curve 
data precludes the application of conventional Fourier methods, and we 
used the algorithm developed by Lomb (\cite{lomb}) following the description by 
Press et al. (\cite{press}).  This algorithm works 
with data $\rm y_i(t_i)$ sampled
at irregular times $\rm t_i$ and normalized by the mean $\rm \bar{y}$ and
gaussian variance $\sigma$.  We calculated these quantities as the 
arithmetic mean and variance of the V magnitudes.  The algorithm then 
compares the normalized data with periodic functions of angular frequency
$\omega$, modeling the data by
$$\rm y_i(t_i) ~ = ~ A ~ cos ~ \omega t_i ~~ + ~~ B ~ sin ~ \omega t_i
\eqno(1)$$
where the $\rm t_i$ are the times at which the observations were made.  The 
Lomb periodogram is then
$$\rm P_N(\omega) ~~ = ~~ {{1}\over{2 \sigma^2}} \left( {{[\sum_i
(y_i ~ - ~ \bar{y}) cos ~ \omega(t_i - \tau)]^2}
\over{\sum_i cos^2 \omega(t_i - \tau)}}\right.$$
$$\rm  + \left.{{[\sum_i (y_i - \bar{y}) sin ~
\omega (t_i - \tau)]^2}\over{\sum_i sin^2 ~ \omega(t_i-\tau)}} \right)
 \eqno(2)$$
where $\tau$, the normalized phase, is given by
$$\rm tan ~ {2 \omega \tau} ~ = ~ {{\sum_i sin ~ 2 \omega t_i} \over
{\sum_i cos ~ 2 \omega t_i}} \eqno(3)$$
The periodogram is computed over a wide range of $\omega$, with 
periodicities showing up as peaks in $\rm P_N(\omega)$.

\begin{figure}
\resizebox{\hsize}{!}{\includegraphics{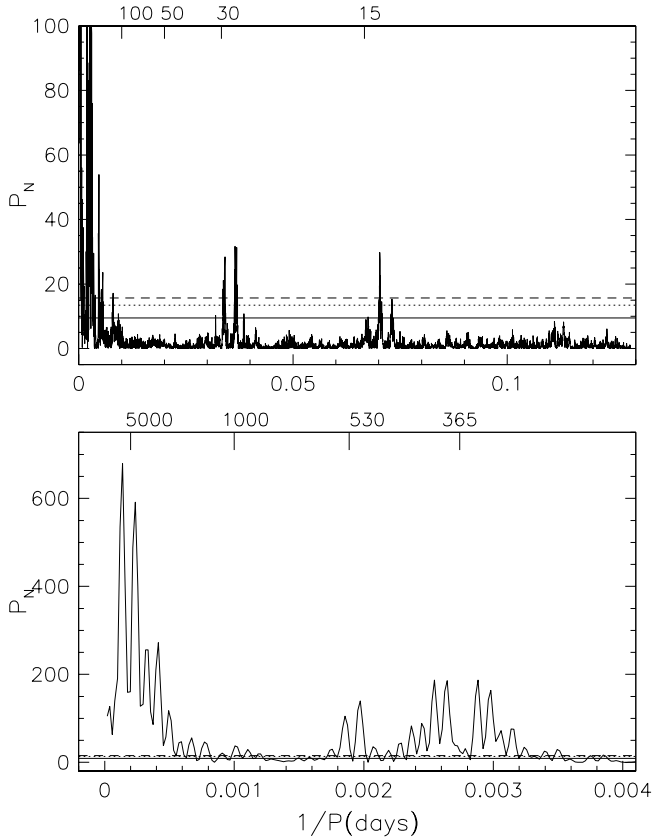}}
\caption{Lomb (\cite{lomb}) periodogram of the optical data for V Hya.
{\bf (a)} Upper panel: entire observed frequency range.  
{\bf (b)} Lower panel:
low frequency range (to periods of about $\rm 250^d$).  The ordinate
is correlation power $\rm P_N$ (see text), the abscissa is frequency in
$\rm days^{-1}$. The period in days is indicated on the upper x-axis.  The
horizontal lines correspond to confidence intervals of 50\% (solid),
99\% (dotted) and 99.9\% (dashed).}
\label{lomb}
\end{figure}

Figure \ref{lomb} shows the periodogram for the 35 years of data for V Hya, 
plotted as power versus frequency f measured in $\rm days^{-1}$.
The upper panel shows the periodogram calculated to the Nyquist frequency,
0.13 $\rm days^{-1}$, and the lower panel the structure of the low-frequency
part of the curve, from periods of infinity to 250 days.  Also plotted
in Fig.\ref{lomb} 
are the approximate 50\%, 99\% and 99.9\% significance levels, 
calculated as described by Scargle (\cite{scargle}) 
and Press et al. (\cite{press}):
$$\rm P(> P_N) ~ \simeq ~ M e^{-P_n}  \eqno(4)$$
where M is taken as twice the number of data points.  

The entire periodogram (Fig.\ref{lomb}, upper panel)
shows highly significant peaks at
$\rm > 300^d$ and approximately $\rm 195^d$, $\rm 29^d$ (this feature
is split into two peaks at about $\rm 28^d$ and $\rm 29^d$), and 
$\rm 15^d$. 
The two peaks at $\rm 28^d$ and $\rm 29^d$
are split by one day, and the two low-significance peaks on either side of
the $\rm 15^d$ peak also correspond to $\pm$1 day.

The lower-frequency region of the periodogram (Fig.\ref{lomb}, lower
panel) shows three
significant peaks (considering the peaks around f = 0.0028 $\rm days^{-1}$
to be part of the same feature).  All three features are broadened,
and give $\rm P_1 ~ = ~ 6600^d \pm 1400^d$, $\rm P_2 ~ = ~ 520^d \pm
30^d$, and $\rm P_3 ~ = ~ 360^d \pm 30^d$.  The approximate 
separation of the peaks near f = 0.0028 $\rm days^{-1}$ corresponds
to about 35 days.  The estimated uncertainties of $\rm P_2$ and $\rm P_3$,
and the substructure in $\rm P_3$, are all about 1 month.

Thus, the data show signatures of all the periods present in the data-taking
process: one year, six months, one month, two weeks and one day.  Even though
the last is outside the frequency range sampled by the data, the irregular
sampling ensures partial sampling of this and even shorter periods.  We
conclude that none of these periods is intrinsic to V Hya, and that the
data support only two stellar periods: $\rm P_1 ~ = ~ 6600^d$ and $\rm P_2
~ = ~ 530^d$.  The large uncertainty in the longer period is
due both to the small number of cycles which have been sampled in the 
data and to the presence of the 
shorter-period variations, both intrinsic and observational. 

To investigate this further, we analyzed 
the time series directly, using the data-folding
technique of Schwarzenberg-Czerny (\cite{schwarz}) who describes the use of 
one-way analysis of variance for searching for and measuring the periods of
variable stars.  
The basis of this method is
to fold the observed light curve $\rm y_i(t_i)$ with some trial period
P, i.e. to cut the data sequence into portions P long and examine the
normalized point-to-point variance of the folded data.  The scatter around 
the average value of the data $\rm \bar{y}$ will be close to random regardless
of the value of P  unless the data stream has some periodicity close to P.
The scatter of the folded data about $\rm \bar{y}$ is then computed for
each value of P and compared with random scatter.

If N is the number of observations $\rm y_i(t_i)$, 
$$\rm \bar{y} ~~ = ~~ {{\sum_{i=1}^N y_i(t_i)}\over{N}} \eqno(5)$$
The data are now folded and binned into M subsets with $\rm N_k$ points in the
{\it kth} bin.  Define $\rm y_{ij}$ as the {\it jth} point in the 
{\it ith} bin, and 
$$\rm \theta ~ = ~ {{S_1^2}\over{S_2^2}} \eqno(6)$$
where
$$\rm S_1^2 ~ = ~ {{\sum_{i=1}^N N_i (\bar{y_i} - \bar{y})^2}
\over {M - 1}} \eqno(7)$$
and
$$\rm S_2^2 ~ = ~ {{\sum_{i=1}^M \sum_{j=1}^{N_i} (y_{ij} - \bar{y})^2}
\over{N - M}}  \eqno(8)$$
$\theta$, the test statistic for the analysis of variance method, equals
1 for pure gaussian noise and is greater than 1 if there is any non-
random component in the data.  The stellar period is found by plotting
$\theta$--1 versus P; values of P for which $\theta$--1 is larger than some 
threshold, corresponding to some confidence level, are considered to be 
periods of the star.  

The analysis of variance code written by A. Udalski (Udalski
et al. 1994) was used to analyze the
light curve of V Hya shown in Fig.\ref{lightcurve}.  Two periods at $>$ 99\% 
significance are found: $\rm P_1 = 6160^d \pm 400^d$, and $\rm P_2
~ = ~ 529.4^d \pm 30^d$ (these are final values - see below).  
Several spurious periods, at a significance level 
below 80\%, were found: $\rm P_3 ~ = ~ 2 P_2$; $\rm P_4 ~ = ~ 2/3 P_2
~ = ~ 344^d$; and $\rm P_5 ~ = ~ 2 P_4$.

The data for V Hya have several properties which cause problems for this 
analysis.  The time interval spanned by the data is
only 2.4 $\rm P_1$, and the magnitude estimates in 
the deep stellar minimum are often poorly determined and/or lower limits.  
The third problem is that the amplitudes of the variations are different, which
increases the range of the data and hence the ``noise'' for individual
period determinations.  We therefore
refined our estimates of the periods of V Hya as follows.  First, 
initial estimates of the
amplitude, period and phase for the long-term variation
($6160^d$) were determined from the folded data.  This variation
was approximated by a sine wave and subtracted.
The residual data were then folded to redetermine the  periods.
The light curve for $\rm P_2$ was determined, and subtracted.  No
significant further periods were found in the residual data.  The 
amplitude, period and phase of maximum light for $\rm P_1$ and $\rm
P_2$ are given in Table 1.

\begin{figure}
\resizebox{\hsize}{!}{\includegraphics{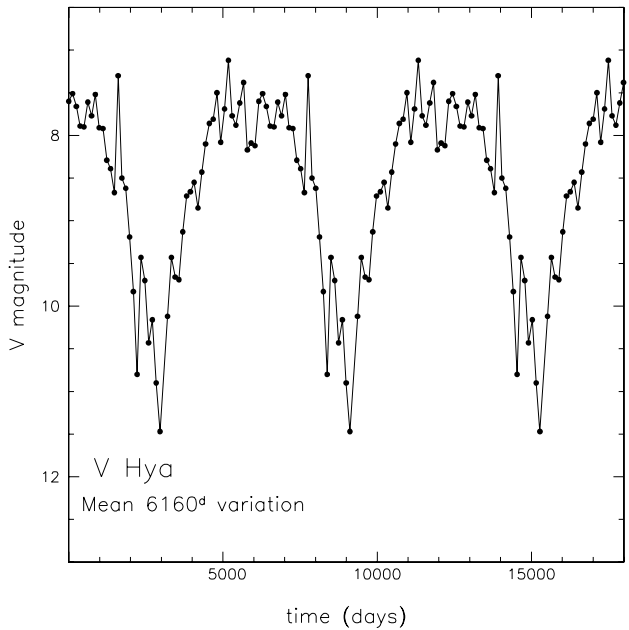}}
\caption{Mean light curve for the $\rm 6160^d$ period of V Hya.
The short-period variation has been approximately subtracted from the 
data, which were then folded to the $\rm 6160^d$ period.  The data are
binned to give 6160 points per cycle, and the mean cycle is repeated several
times.}
\label{p6160}
\end{figure}

\begin{figure}
\resizebox{\hsize}{!}{\includegraphics{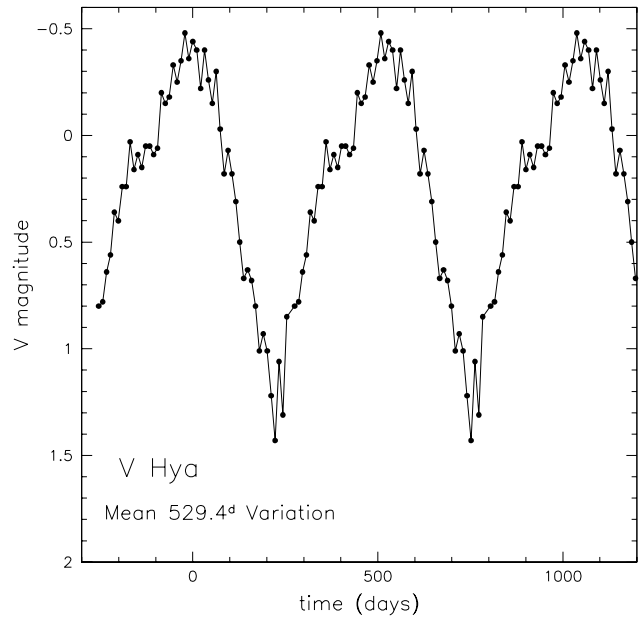}}
\caption{Mean light curve for the $\rm 529^d$ variation of V Hya.
The longer-period variation has been approximately subtracted from the data,
which were then folded to the $\rm 529^d$ period.  The data are binned to
give 50 points per cycle, and the mean cycle is repeated several times.}
\label{p529}
\end{figure}

The resulting folded light curves, for $\rm P_1 ~ = ~ 6160^d$ with 
the $\rm 529.4^d$ variation subtracted and for $\rm P_2 ~ = ~ 529.4^d$
with the $\rm 6160^d$ variation subtracted, are shown in Figs.
\ref{p6160} and~\ref{p529}.
The plotted light curves are averaged into 50 bins per cycle and the cycle 
repeated several times.  As discussed above, the minima are not well
determined.  Fig.\ref{p529} shows a fairly typical Mira-type light curve, 
to first order a sine-wave variation but with a slow rise and more 
rapid decline.  The long-period variation (Fig.\ref{p6160}) is not approximately
sinusoidal however; it resembles the light curve of an eclipsing binary
star, but with a far longer period and duration.  The data of Mayall
(\cite{mayall}) show a similar light curve shape, 
and cover five previous long-period
cycles, with $\rm P_1 ~ = ~ 
6200^d \pm 400^d$, in good
agreement with the more recent data.  The long-period variation in V Hya
thus appears to be very regular. 

Finally, we note that the amplitude of the 17-year variation, $\rm 3.5^m$,
agrees with that given by Kholopov et al. (1985) and is
smaller than the amplitude of $\rm 5^m - 6^m$ suggested by Mayall, 
which is actually
the total range of variation due to both periodicities.

What is the origin of the $\rm 6000^d$ variation of V Hya?
The great depth of the minimum, and V Hya's large mass loss rate
(Paper I), suggest obscuration by dust, analogous to the ejection
of obscuring material by R CrB stars; however, the dimming of R
CrB stars is irregular.  The regularity of V Hya's $\rm 6000^d$
variation support, rather, a dynamical origin for the variation.
We suggest that V Hya is an eclipsing binary, but that the 
eclipse is caused not by a stellar companion but by circumstellar dust.
Similar phenomena are seen in a small number of
stars of widely different spectral types and at very
different evolutionary stages. One such star is the F0 supergiant 
$\epsilon$ Aur, which undergoes an eclipse every 27.1 years.
There is no secondary
eclipse, and the primary eclipse is of long duration, (22 months)
showing that the secondary cannot be a star but must be a large (5 - 10 A.U.)
dark body, which observations strongly suggest is a dust disk 
orbiting with a companion (Huang \cite{huang}; McRobert \cite{mcrobert85},
\cite{mcrobert88}; 
Lissauer et al. \cite{lissauer}). 
Eclipses best explained by dust clouds attached to a binary companion 
are also seen in a small number of planetary nebula nuclei, e.g. in NGC 2346
(Costero et al. \cite{costero}).  The shape of V Hya's light curve 
(Fig.\ref{p6160})
suggests that it may be a similar case. 
V Hya's period of $\rm 6160^d$, if it is identified as an orbital period, 
corresponds to a distance from the star
of $\rm \sim 3.5 \times 10^{14}$ cm (about
10 $\rm R_{\star}$) and a circular velocity of 2 $\rm km~s^{-1}$
(assuming the mass of the system is 1 $\rm M_{\odot}$).  The transit time
for an orbiting body is then about $\rm 2200^d$ and its diameter
at least several $\rm \times 10^{13}$ cm.  
V Hya is a very evolved AGB star losing mass at a high rate,
and the circumstellar envelope is flattened (Kahane et al. \cite{kahane96}); 
this could be caused by a
companion orbiting in the inner parts of the
circumstellar dust shell
(e.g. Mastrodemos \& Morris \cite{mastrodemos}). 
If this is taking place in V Hya's envelope, the 
eclipses of the star may be caused by a flattened
density enhancement or wake in
the envelope due to an orbiting companion.

\section{Discussion and Conclusions}

\subsection{The Evolutionary State of V Hya}

V Hya is a very unusual star; its $\rm 530^d$ and $\rm 6000^d$
periods are unique among evolved stars, its mass loss rate is
high, and it is ejecting a very fast molecular wind (Paper I, 
Appendix A).  Its molecular line envelope shows kinematic structure
which has been interpreted as a bipolar outflow (Tsuji et al.
\cite{tsuji}, Kahane et al. \cite{kahane88}, \cite{kahane96})
or as a tilted, expanding disk (Paper I).  The optical spectrum
shows unusually broadened lines, suggesting that the star is 
rapidly rotating (Barnbaum et al. \cite{barnbaum}, 
hereafter BMK).  Although some of these
properties suggest that the star may be evolving beyond the AGB
phase, its spectral type, IRAS colors and lack of radio continuum
emission (Paper I, Appendix A) are all consistent with its being an 
AGB star.

BMK show that V Hya's rotation velocity is period
dependent, decreasing as the star expands, 
and suggest that the rapid rotation is due to the star
having evolved to a common envelope binary; the companion may, also,
be responsible for the enhanced C abundance in the envelope (V Hya
does not show the presence of Tc).  The inferred rotation velocity
suggests a period about 10\% longer than the pulsation period of the
star, with a large uncertainty due to lack of knowledge of the
inclination: BMK suggest that this similarity in the pulsation and
rotation periods may be responsible for the $\rm 6000^d$ variation,
which is identified as a beat frequency of the system.  Further, they suggest
that the rapid rotation of the star can drive the fast outflow.  

If the interpretation of V Hya's light curve given in the previous
section of the present paper
is correct, on the other hand,
only the $\rm 530^d$ variation is intrinsic
to the star, and the amplitude and period of this variation
suggest that V Hya is a normal Mira.  The suggested eclipse period
of 17 years implies an external companion, but one which is close
enough to the star to shape its outflowing wind into a flattened
configuration (cf. Mastrodemos and Morris 1998) and, perhaps, to
cause the fast wind.  Of particular interest for both of these
scenarios is the high mass loss rate ($\rm 4 \times 10^{-5}
M_{\odot}
yr^{-1}$) and small envelope size inferred from low resolution CO
mapping observations (Paper I, Appendix A).  The latter shows that
the duration of the very high mass loss rate phase has been quite
short,$<$ 2000 years.  The envelope mass, about 0.1 $\rm M_{\odot}$,
is similar to that found for many planetary nebulae. The recent 
onset of copious mass loss could be due to the formation of a
common-envelope binary, as suggested by BMK, or could be due to
the formation of the superwind which signals the end of evolution
on the AGB.  Whatever the reason(s) for the peculiarities of V Hya,
the observational data show that the complex structures seen
in many planetary nebulae (axisymmetric structure, fast winds)
are often present in the envelope of the precursor star before it evolves
beyond the AGB.

\subsection{Conclusions}

This paper discusses several new observations of the evolved AGB carbon
star V Hya. We find:

\begin{enumerate}

\item An analysis of optical variability data finds only two significant
periods in V Hya, at $\rm 529.4^d$ with peak-to-peak variation of
$\rm 1.5^m$, and $\rm 6160^d$, with peak-to-peak variation of $\rm
3.5^m$.  Both of these
periods are seen in data taken back to 1880, and both appear to be 
very regular.

\item
The morphology of the $\rm 6160^d$ variation resembles that of an eclipsing
binary, but with an eclipse duration which is far longer than can
be produced by a stellar companion and of an amplitude which shows that
essentially the entire stellar photosphere is occulted.  We suggest
that the regular long-period dimming of V Hya is due to a thick dust cloud 
orbiting the star and attached to a binary companion.  

\item The $\rm 529.4^d$ period is typical of a luminous Mira variable,
and we suggest that V Hya is a Mira, not a semiregular
variable.  The persistence of this variation to the present day is
further evidence that V Hya is still on the AGB. 

\item No radio frequency emission was detected from V Hya to  a
limit of S(8.4 GHz) $<$ 0.17 mJy.  This shows that the star
is not  hot enough to cause ionization of the
envelope and is thus not yet evolved away from the
AGB towards the planetary nebula phase.  

\item Emission from the fast molecular wind was observed in the submillimeter
CO(4--3) and CO(6--5) lines.  Together with the CO(2--1) and CO(3--2)
observations of Paper I, the relative line intensities
show that both the fast wind and the
slower wind appear to be of recent origin, only $\sim$ 1500 years in
the case of the slower wind, and that copious mass loss 
(at rates of several $\rm \times
10^{-5} ~ M_{\odot} yr^{-1}$) began only recently.  The gas in the fast
wind does not appear to be significantly hotter than that in the slow 
wind, arguing against formation of the molecules in dense hot postshock
gas and in favor of the direct production of the fast wind by V Hya.

\item The presence of the fast molecular wind and the recent
onset of rapid mass loss suggest that V Hya may
be in the initial stages of its evolution beyond the AGB.  The formation
of fast winds seems often to accompany the post-AGB evolution of a star.
These observations of V Hya show that fast
molecular winds and complex envelope structure
can form while the star is still on the AGB.
\end{enumerate}

\begin{acknowledgements}

We wish to thank the referee for a most thorough and helpful
report on the first draft of this paper.
We are very grateful to the staff at the CSO and VLA for granting
the observing time used for part of this work, and to Norbert Bartel
for donating an hour of VLA time during a VLBI run. This research made use
of the SIMBAD data base, operated at CDS, Strasbourg, France.  We thank
Bohdan Paczynski, Przemyslaw Wozniak and Andrzej Udalski for advice
and for providing the analysis-of-variance code.  Astronomical research at
the CSO is supported by the National Science Foundation via grant 
AST96-15025.  Partial support for this work from Princeton University
and from the N.S.F. via grant AST96-18503 is gratefully acknowledged.
\end{acknowledgements}

\appendix
\section{Observations of V Hya and its Circumstellar
Envelope}

Luttermoser \& Brown (\cite{lb}) detected
0.22$\pm$0.02 mJy
at 8.4 GHz from V Hya using the Very Large Array (VLA).  
This is far stronger 
than expected for either the circumstellar dust or the stellar photosphere,
and Luttermoser and Brown postulate
an extended, partly ionized atmosphere.
Another possible source of thermal radio frequency emission from circumstellar
envelopes is ionization produced as the star
evolves away from the AGB.
Since evolution is rapid at this stage, the increase in the
strength of the continuum emission from the compact central HII region can
be observed in only a few years, as is found for AFGL 618 (Kwok \&
Feldman \cite{kwok}).  Since V Hya may be in the initial stages of evolution 
away from the AGB, we re-observed the star with the VLA to investigate 
whether the radio continuum flux density has changed.

V Hya was observed at 3.6 cm (8.4 GHz)
on Dec 13, 1996.  The array
was in its `A' (largest spacing) configuration.  Two short observations,
totaling 20 minutes of observing time, were made.  The phases were calibrated
by short observations of the phase calibrator 1048-191 before and after
each set of observations.  The data were of excellent quality.  The flux 
density scale was calibrated relative to 3C 48 ($\rm S_{8.4}$ = 3.2424 Jy)
and 3C 286 ($\rm S_{8.4}$ = 5.2001 Jy) (Taylor \& Perley \cite{taylor}).
No emission was seen at
or near the position of V Hya, with a
3$\sigma$ upper limit of 0.17 mJy.
Thus, the star has not yet 
evolved away from the AGB to the point where it is hot enough
to produce ionizing photons.

\begin{figure}
\resizebox{\hsize}{!}{\includegraphics{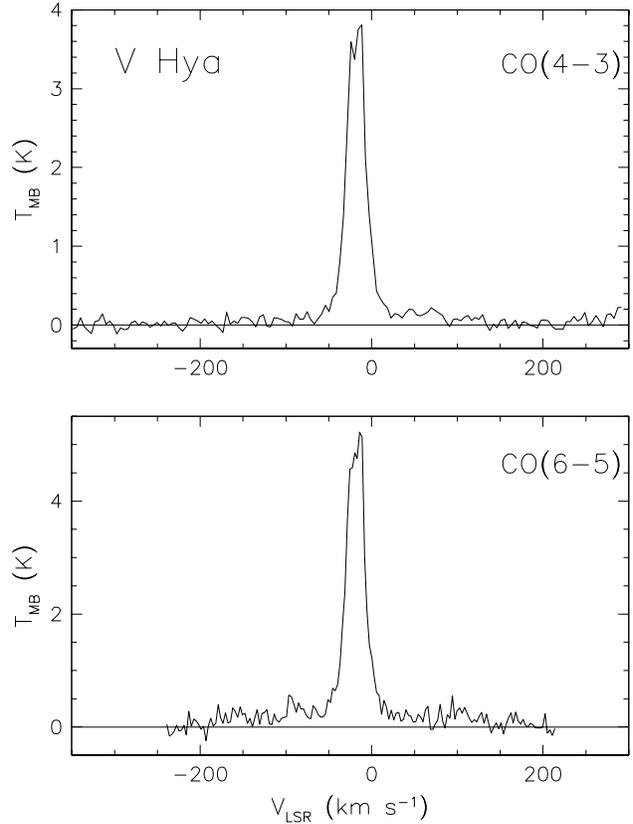}}

\caption{Submillimeter CO line emission from V Hya observed at
the CSO.  (a) Upper panel: CO(4--3) line at 650$\rm \mu m$ and (b) lower panel:
CO(6--5) line at 440$\rm \mu m$.  The ordinate is main-beam brightness 
temperature and the abscissa velocity with respect to the Local Standard of
Rest. Both line profiles have had a linear baseline removed.  The fast 
wind is detected at the 2$\sigma$ level in the CO(4--3) line and the 
3$\sigma$ level in the CO(6--5) line.}
\label{co}
\end{figure}

\begin{table}
\caption[]{CO Emission from V Hya's Circumstellar Envelope}
\vspace{0.5cm}
\begin{tabular}{lllll}
  line &
  T(wing)&
  T(peak)& 
  T(wing)/T(peak) & 
  $\theta$ \cr
  &
  (K)&
  (K)&
  &
  ($''$)\cr
\vspace{0.5cm}

CO(2-1)& 0.044& 1.75& 0.025& 30\cr
CO(3-2)& 0.104& 3.30& 0.032& 20\cr
CO(4-3)& 0.127& 3.80& 0.034& 15\cr
CO(6-5)& 0.188& 5.21& 0.036& 10\cr
\end{tabular}
\end{table}

CO observations of V Hya's envelope
were made on the nights of March 24-28 with 
the 10.4 m Robert B. Leighton telescope of the Caltech Submillimeter 
Observatory on Mauna Kea, Hawaii.  The weather was excellent, with a
zenith opacity at 220 GHz of $\rm \tau_{\circ} ~ \leq$ 0.03. 
The CO(4--3) line at 461.0408 GHz and the CO(6--5) line at 691.473 GHz
were observed.  The line profiles
are shown in Fig.\ref{co}
 - both the fast 
and normal winds are seen. Table 2 summarizes the CO line
data obtained at the Caltech Submillimeter Observatory for V Hya, 
giving the brightness temperature of the fast wind
(measured at +100 $\rm km~s^{-1}$), the brightness temperature of the
main component (measured at about --10 $\rm km~s^{-1}$), the ratio
T(fast wind)/T(slow wind), and the half-power beamwidth of the CSO
for each line.  The CO(2--1) and CO(3--2) data are from Paper 1.
These data show that the ratio of the CO(2--1) and
CO(3--2) line intensities for both the fast and slow winds have
about the value expected for an optically thick source which is
roughly the same size as or smaller than the $\rm 20''$ beam for the
CO(3--2) line.  The ratios of the CO(6--5), CO(4--3) and CO(3--2)
lines, however, show that the envelope is partly
resolved at 460 and 691 GHz. The fast wind does not appear to be 
significantly hotter than the slow wind. These observations
confirm the conclusions from Paper I: V Hya is ejecting a very fast 
molecular wind as well as slower winds, and the circumstellar
envelopes formed by both the fast and slow winds have approximately the
same diameter, $\rm 20''$.  These data give a radius of $\rm \sim 
8 \times 10^{16}$ cm at a distance of 500 pc, a dynamical age for the
slow wind of about 1600 years, and a total mass loss rate of about $\rm 4
\times 10^{-5} ~ M_{\odot} ~ yr^{-1}$. This high mass loss rate, and the
small dynamical age of the envelope, suggests that V Hya has entered its 
``superwind'' phase.

\end{document}